# A New Guard-Band Call Admission Control Policy Based on Acceptance Factor for Wireless Cellular Networks

Md. Asadur Rahman, Mohammad Arif Hossain, Shakil Ahmed, and Mostafa Zaman Chowdhury
Department of Electrical and Electronic Engineering
Khulna University of Engineering & Technology, Khulna-9203, Bangladesh
E-mail: eeeasadur@yahoo.com, dihan.kuet@gmail.com, shakileee076@gmail.com, mzceee@yahoo.com

*Abstract*—To ensure the maximum utilization of the limited bandwidth resources and improved quality of service (QoS) is the key issue for wireless communication networks. Excessive call blocking is a constraint to attain the desired QoS. In cellular network, as the traffic arrival rate increases, call blocking probability (CBP) increases considerably. Paying profound concern, we proposed a scheme that reduces the call blocking probability with approximately steady call dropping probability (CDP). Our proposed scheme also introduces the acceptance factor in specific guard channel where originating calls get access according to the acceptance factor. The analytical performance proves better performance than the conventional new-call bounding scheme in case of higher and lower traffic arrival rate.

## I. Introduction

The cellular communication is one of the best techniques in wireless communication systems for an efficient radio resource management due to its high mobility management. Consequently, there increases a great demand for Personal Communication Services (PCS) which will provide reliable communications via lightweight and pocket-size terminals [1]. The base station (BS) of the cellular network is responsible for assigning channel to each call. When a mobile user crosses the cell boundary or the quality of the wireless link is unacceptable, then the process of handoff call is initiated [2]. In recent years, a remarkable tendency in the design of wireless cellular systems is decreasing in the cell size and increasing the user mobility. These two factors result in more frequent handovers in wireless communication system [3].

A call admission control (CAC) scheme aims to maintain the delivered QoS to the different calls at the target level by limiting the number of enduring calls in the system. One major challenge in designing a CAC arises to provide service two major types of calls: new calls or originating calls and handoff calls. The QoS performances related to these two types of calls are generally measured by new call blocking probability and handoff call dropping probability. In general, users are more sensitive to dropping of an ongoing and handed over call than blocking a new call [4].

Every CAC scheme has certain constraint to reduce the network blockage and termination of new calls and handoff calls. In [1]-[8], some CAC schemes have been proposed. Since blocking a new call is less serious than dropping a handoff call, CAC schemes usually give a higher priority to handoff calls. Various one dimensional handoff priority-based CAC schemes have been proposed in [1], [6]-[8]. In these CAC schemes, there is a tradeoff between handoff calls and new calls. That means due to provide priority to the handoff calls the blocking probability of new calls increased.

In this paper, we propose a new guard-band CAC scheme based on new-call bounding scheme [6] and acceptance probability of call arrival rate. By this scheme we have shown that in a specific new call-bounding scheme blocking probability can be reduced without changing the handoff call dropping probability. The novelty of the acceptance factor is to determine the lower blocking probability whether the traffic arrival rate is less or more. We also describe the impact of this acceptance factor on blocking probability. Else the various performances of this scheme on different conditions are analyzed. The behavior of the proposed scheme is studied using one dimensional Markov chain and we present some uniqueness of this CAC scheme.

This paper is organized as follows: Section II shows the hypothesis on handover and new call ratio. The new-call bounding scheme is explained in Section III. In Section IV we represent our proposed scheme. The performance of this paper is analyzed in Section V. Finally conclusion about the total work is drawn in Section VI.

## II. Hypothesis on Handover & New Call Ratio

In macro-cellular networks, the rate of new-call and handover call does not maintain the fixed ratio. This is why a hypothesis is necessary to obtain the relation between them. The relation among the originating or new call arrival rate ($\lambda_n$), the handoff call arrival rate ($\lambda_h$) and the average channel departure rate ($\mu$) is essential to determine the call blocking and dropping the handoff call request. Here, it is considered that $P_B$ and $P_D$ will represent the blocking probability of new calls and the dropping probability of handoff calls request respectively. All call arriving processes are assumed to be as Poisson's distributed.

A new call that arrives in the system may be either completed within the original cell or handover to another cell or cells before completion. The probability of handover of a call depends on two factors, (i) the average dwell time ($1/\eta$) (ii) the average call duration ($1/\mu_a$) [4]. Again the average channel departure rate ($\mu$) also depends on the above two parameters. Since both the call duration and the cell dwell time are assumed to be exponential, the handover probability, $P_h$ of a call at a particular time is given by [10]:

$$P_h = \frac{\eta}{\eta + \mu_a} \quad (1)$$

and the handoff call arrival rate into a cell is evaluated as:

$$\lambda_h = \frac{(1-P_B)P_h}{[1-P_h(1-P_D)]} \quad (2)$$

where the equation agrees from balancing the rates of handover calls into and out of a cell.

When a call is originated in a cell and gets a channel, the call holds the channel until the call is completed in the cell or the mobile moves out of the cell. Therefore, the channel holding time $T_C$ is either dwell time, $T_h$ or the call length time, $T_n$ [10]. Then the relation among them can be represented as following below.

$$T_c = \min(T_h, T_n) \quad (3)$$

## III. NEW-CALL BOUNDING SCHEME

New-call bounding scheme is a general priority scheme. In this case, priority is given to handoff requests by assigning guard channels ($G_C$) entirely for handoff calls among the $C$ channels in a cell. The rest $M (= C - G_C)$ channels are shared by both new calls and handoff calls [9]. A new call is blocked if the number of available channels in the cell is less than or equal to $M$. A handoff request is blocked if no channel is accessible in the target cell.

The state $i$ ($i = 0, 1, \ldots C$) of a cell is defined as the number of calls in progress for the BS of that cell. Let $P(i)$ be the steady-state probability that the BS is in state $i$. The probabilities $P(i)$ can be found by the typical way of birth–death processes. The relevant state transition diagram is shown in Figure 1. From the figure, the state balance equations are-

$$i\mu P(i) = \begin{cases} (\lambda_n + \lambda_h)P(i-1) & 0 \leq i \leq M \\ \lambda_h P(i-1) & M < i \leq C \end{cases} \quad (4)$$

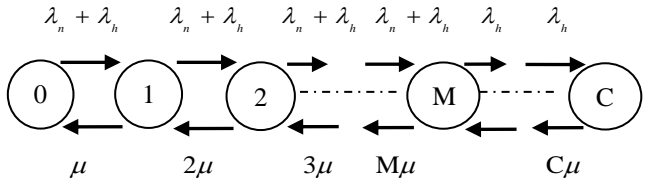

Figure 1: State transition diagram of new-call bounding scheme.

The steady-state probability $P(i)$ is found as follows:

$$P(i) = \begin{cases} \dfrac{(\lambda_n + \lambda_h)^i}{i!\mu^i} P(0) & 0 \leq i \leq M \\ \dfrac{(\lambda_n + \lambda_h)^M \lambda_h^{i-M}}{i!\mu^i} P(0) & M < i \leq C \end{cases} \quad (5)$$

where

$$P(0) = \left[ \sum_{i=0}^{M} \frac{(\lambda_n + \lambda_h)^i}{i!\mu^i} + \sum_{i=M+1}^{C} \frac{(\lambda_n + \lambda_h)^M \lambda_h^{i-M}}{i!\mu^i} \right]^{-1} \quad (6)$$

The blocking probability, $P_B$ for a new call is given by-

$$P_B = \sum_{i=1}^{M} P(i) \quad (7)$$

By this way, the blocking probability of handoff request or dropping probability, $P_D$ is given by-

$$P_D = \frac{(\lambda_n + \lambda_h)^M \lambda_h^{C-M}}{C!\mu^C} P(0) = P(C) \quad (8)$$

## IV. PROPOSED CAC SCHEME

In the proposed scheme, we use the basic idea of new-call bounding scheme and also a special guard band inside the channels that accepts the new calls with a defined acceptance factor and rejects the rest new calls. This guard band is assigned between priority and non-priority band by taking some channels of the guard band from only handoff accessing channels. This is why, in our proposed scheme, priority is given to the handoff-call by two steps. The state transition diagram of the system is described clearly by Markov chain in Figure 2. The three steps of the total channel ($C$) allocation can be categorized briefly as following below.

1. 0~M Channel can be used by handoff and new calls with same acceptance probability

2. M~N channel are allocated for handoff and new calls with specific acceptance probability. Though the new call is accepted with the factor α, the handoff call will be accepted with the probability 1.

3. Rest N~C channels are allocated for handoff request only. Here new call will get the acceptance probability 0, that is, α=0.

The proposed scheme is designed mathematically in such a way that shows the general characteristic of new-call bounding scheme. When the value of acceptance probability of this scheme becomes zero, this scheme turns to be same as new-call bounding scheme. Besides, the value of acceptance factor 1 poses the characteristics of non-priority scheme. Depending on these ideas the total mathematical expression is set.

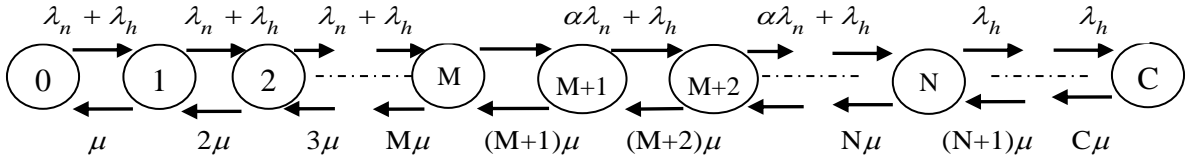

Figure 2: Markov chain diagram of the proposed scheme.

The steady-state probability *P(i)* is easily found as follows:

$$P(i) = \begin{cases} \dfrac{(\lambda_n + \lambda_h)^i}{i!\mu^i} P(0) & 0 \leq i \leq M \\[2mm] \dfrac{(\lambda_n + \lambda_h)^M [\alpha\lambda_n + \lambda_h]^{i-M}}{i!\mu^i} P(0) & M < i \leq N \\[2mm] \dfrac{(\lambda_n + \lambda_h)^M [\alpha\lambda_n + \lambda_h]^{M-N} \lambda_h^{i-N}}{i!\mu^i} P(0) & N < i \leq C \end{cases} \quad (9)$$

where

$$P(0) = \left[ \sum_{i=0}^{M} \frac{(\lambda_n + \lambda_h)^i}{i!\mu^i} + \sum_{i=M+1}^{N} \frac{(\lambda_n + \lambda_h)^M [\alpha\lambda_n + \lambda_h]^{i-M}}{i!\mu^i} + \sum_{i=N+1}^{C} \frac{(\lambda_n + \lambda_h)^M [\alpha\lambda_n + \lambda_h]^{N-M} \lambda_h^{i-N}}{i!\mu^i} \right]^{-1} \quad (10)$$

The blocking probability $P_B$ for a new call according to (9)-(10) is given by-

$$P_B = \frac{(\lambda_n + \lambda_h)^M (1-\alpha)(\alpha\lambda_n + \lambda_h) P(0)}{\mu^{M+1}} \times \sum_{i=1}^{N-M} \frac{[\alpha\lambda_n + \lambda_h]^{i-1}}{(M+1)!\mu^i} + P(0) \sum_{i=N+1}^{C} P(i) \quad (11)$$

The dropping probability for handoff request will be same as the new call bounding scheme,

$$P_D = P(C) \quad (12)$$

From (9), (10) and (11), it is clear that if we put the acceptance factor, $\alpha = 0$ the proposed scheme becomes the new-call bounding scheme or conventional guard band policy for handover call priority scheme. Else, if we put the acceptance factor, $\alpha = 1$ the scheme turns to be non-priority scheme. This is the generalization of new-call bounding scheme. According to the value of acceptance factor the blocking probability will fluctuate in nonlinear pattern about which we will explain in our next section.

## V. Performance Analysis

The proposed scheme and the conventional new call bounding scheme are analyzed with average call life time $1/\mu_a$=120 second and average cell dwell time $1/\eta$=360 second. Total number of channel, *C* in both case is taken as 130.

Handoff and new call equally sharing channel number, *M* is taken 100. According to the proposed scheme the special guard band with acceptance factor is defined by 10 channels which is allocated between *M* and *N*, this is why *N*=110.

In our proposed scheme, it is necessary to find the value of acceptance factor that shows the minimum blocking and dropping probability. In this case, we have analyzed the value of acceptance factor by iterative method from 0.1 to 0.9. We have found that the value of acceptance factor *(α)* that demonstrates minimum blocking probability and steady dropping probability throughout the call arrival rates is 0.9 for lower traffic load and at the higher traffic load the acceptance factor 0.5 shows the minimum blocking probability.

By this consideration, taking $\alpha = 0$ the blocking probability of new calls and the dropping probability of handoff-calls for the proposed scheme and these probability for new-call bounding scheme have been shown in Figure 3.

In Figure 4, we present a comparison among the call blocking probability of different acceptance factor. Form the figure it has been observed that at higher traffic rate the call blocking probability of acceptance factor 0.5 shows the minimum level. But for the lower traffic rate the factor 0.9 shows the minimum blocking probability. But in every case the dropping probability of handoff calls are considerably constant which is shown in Figure 5.

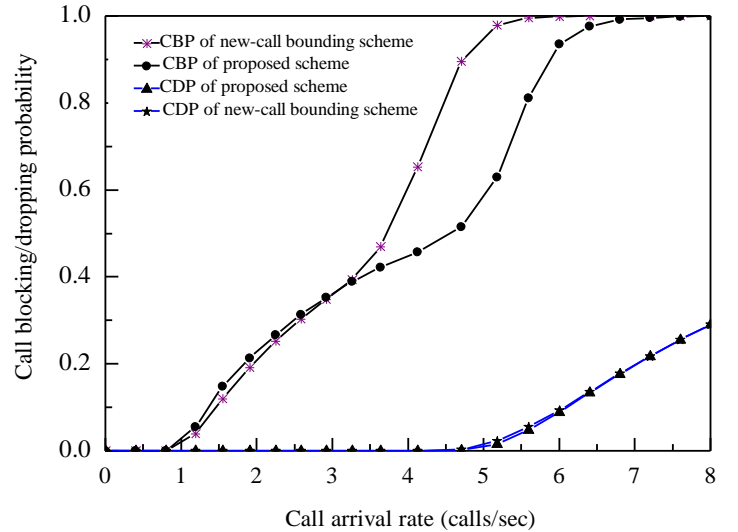

Figure 3: Comparison of call blocking and call dropping probability between new-call bounding scheme and proposed scheme.

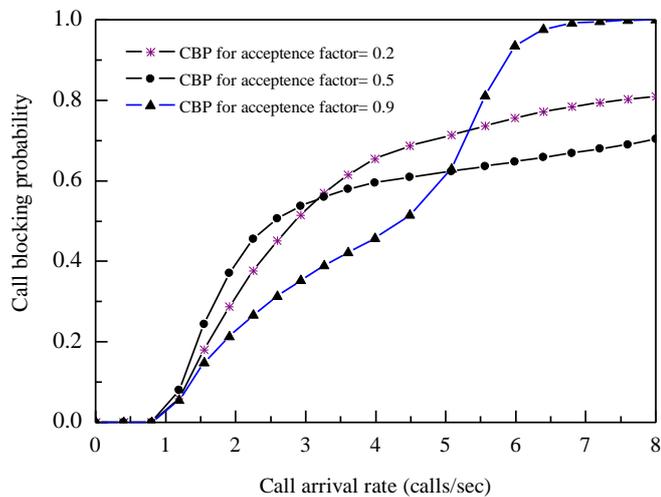

Figure 4: Comparison of call blocking probability with different acceptance factor.

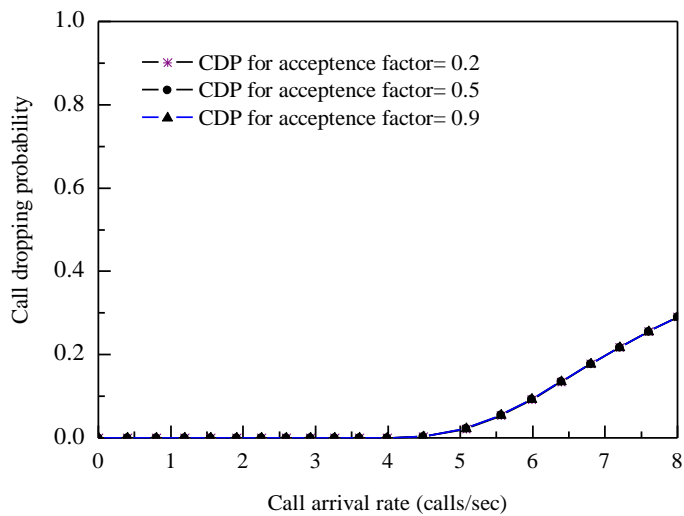

Figure 5: Comparison of call dropping probability with different acceptance factor.

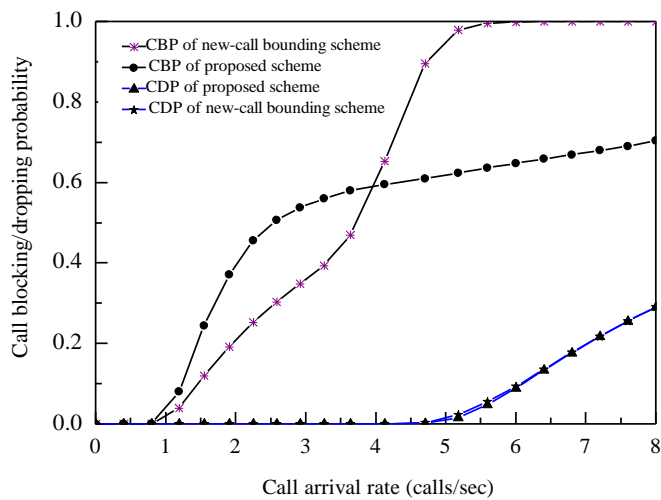

Figure 6: Comparison of proposed scheme and new-call bounding scheme for acceptance ratio of 0.5.

As we said that the value of acceptance factor 0.5 shows the minimum blocking probability at the higher traffic load with respect to that of new-call bounding scheme. In Figure 6 we illustrated that graphical presentation.

## VI. CONCLUSIONS

In this paper, an efficient guard-band CAC scheme has been proposed which combines the idea of new-call bounding scheme and call acceptance dependent CAC scheme. From the derived equations of the proposed scheme we can return back to new-call bounding CAC scheme. This proposed CAC scheme ensures a minimum permissible call blocking probability keeping the call dropping probability almost constant as new-call bounding scheme. In the new-call bounding scheme there is a fixed guard band that provides lower performance but our proposed scheme contains an extra narrow guard band with acceptance factor that ensures better performance. Else this work clarifies to choose the value of acceptance factor whether the traffic arrival rate is lower or higher. In our future work, we will research on CAC policy using two dimensional Markov chain for multiclass traffic.

REFERENCES

[1] R. Ramajee, R. Nagarajan, and D. Towsley, "On optimal call admission control in cellular networks," *Wireless Networks*, vol.3, no.1, pp.29-41, 1997.

[2] A. Sgora, Vergados, and Dimitrios D., "Handoff prioritization and decision schemes in wireless cellular networks: a survey," *IEEE Communications Surveys and Tutorials*, vol.11, no.4, pp.57-77, December 2009.

[3] M. Z. Chowdhury, Y. Min Jang, and Z. J. Haas, "Call admission control based on adaptive bandwidth allocation forwireless networks," *Journal of Communications and Networks*, vol.15, no.1, pp.15-24, February 2013.

[4] A. Leelavathi and G. V. Sridhar, "Adaptive bandwidth allocation in wireless networks with multiple degradable quality of service," IOSR *Journal of Electronics and Communication Engineering*, vol.2, no.4, pp. 25-29, October 2012.

[5] Y. Fang, "Thinning scheme for call admission control in wireless networks," *IEEE Transactions on Computers*, vol.52, no.5, pp.685-687, May 2003.

[6] Y. Fang and Y. Zhang, "Call admission control schemes and performance analysis in wireless mobile networks," *IEEE Transactions on Vehicular Technology*, vol. 51, no. 2, pp.371-382, March 2002.

[7] D. Hong and S. S. Rappaport, "Traffic model and performance analysis for cellular mobile radio telephone systems with prioritized and noprioritized handoff procedures," *IEEE Transactions on Vehicular Technology*, vol.35, no.3, pp. 77-92, August 1986.

[8] J. L. Vazquez Avila, F. A. Cruz Perez, and L. O. Guerrero, "Performance analysis of fractional guard channel policies in mobile cellular networks," *IEEE Transactions on Wireless Communications*, vol.5, no.2, pp.301- 305, February 2006.

[9] Q. An Zeng and D. P. Agrawal, "Handoff in wireless mobile networks," Department of Electrical Engineering and Computer Science, University of Cincinnati, by John Wiley & Sons, Inc, 2002.

[10] M. Schwartz, Mobile wireless communications. Cambridge University Press, 2005.